\documentclass[12pt]{article}

\makeatletter
\renewcommand\section{\@startsection {section}{1}{\z@}%
                                   {-3.5ex \@plus -1ex \@minus -.2ex}
                                   {2.3ex \@plus.2ex}%
                                   {\normalfont\large\bfseries}}
\renewcommand\subsection{\@startsection{subsection}{2}{\z@}%
                                     {-3.25ex\@plus -1ex \@minus -.2ex}%
                                     {1.5ex \@plus .2ex}%
                                     {\normalfont\bfseries}}
\makeatother

\def\IZ{\relax\ifmmode\mathchoice
{\hbox{\cmss Z\kern-.4em Z}}{\hbox{\cmss Z\kern-.4em Z}}
{\lower.9pt\hbox{\cmsss Z\kern-.4em Z}} {\lower1.2pt\hbox{\cmsss
Z\kern-.4em Z}}\else{\cmss Z\kern-.4em Z}\fi}
\def\IR{\relax{\rm I\kern-.18em R}}

\def\one{{\hbox{ 1\kern-.8mm l}}}

\def\tr{{\rm tr\,}}

\newlength{\bredde}
\def\slash#1{\settowidth{\bredde}{$#1$}\ifmmode\,\raisebox{.15ex}{/}
\hspace*{-\bredde} #1\else$\,\raisebox{.15ex}{/}\hspace*{-\bredde}
#1$\fi}

\newsavebox{\zzzbar}
\sbox{\zzzbar}
  {\setlength{\unitlength}{0.9em}
  \begin{picture}(0.6,0.7)
  \thinlines
  \put(0,0){\line(1,0){0.6}}
  \put(0,0.75){\line(1,0){0.575}}
  \multiput(0,0)(0.0125,0.025){30}{\rule{0.3pt}{0.3pt}}
  \multiput(0.2,0)(0.0125,0.025){30}{\rule{0.3pt}{0.3pt}}
  \put(0,0.75){\line(0,-1){0.15}}
  \put(0.015,0.75){\line(0,-1){0.1}}
  \put(0.03,0.75){\line(0,-1){0.075}}
  \put(0.045,0.75){\line(0,-1){0.05}}
  \put(0.05,0.75){\line(0,-1){0.025}}
  \put(0.6,0){\line(0,1){0.15}}
  \put(0.585,0){\line(0,1){0.1}}
  \put(0.57,0){\line(0,1){0.075}}
  \put(0.555,0){\line(0,1){0.05}}
  \put(0.55,0){\line(0,1){0.025}}
  \end{picture}}
\newcommand{\Tr}{\mathop{\mbox{Tr}}\nolimits}

\newcommand{\ena}{\end{eqnarray}}
\newcommand{\beqa}{\begin{eqnarray}}
\newcommand{\eeqa}{\end{eqnarray}}
\newcommand{\bea}{\begin{eqnarray}}
\newcommand{\eea}{\end{eqnarray}}

\newcommand{\be}{\begin{equation}}
\newcommand{\ee}{\end{equation}}

\usepackage{graphicx}

\newcommand{\ex}[1]{\mbox{e}^{\,\textstyle#1}}
\newcommand{\beq}{\begin{equation}}
\newcommand{\eeq}{\end{equation}}
\newcommand{\ber}{\begin{array}}
\newcommand{\eer}{\end{array}}

\newcommand{\dsty}{\displaystyle}

\newcommand{\de}{\delta}

\newcommand{\cnst}{\mbox{const}}

\newcommand{\eps}{\varepsilon}

\usepackage{amsmath}
\usepackage{amssymb}

\begin{document}
\begin{titlepage}
\begin{flushright}
\phantom{arXiv:yymm.nnnn}
\end{flushright}
\begin{center}
{\Large\bf Light-like Big Bang singularities\vspace{2mm}\\ in string and matrix theories}    \\
\vskip 10mm
{\large Ben Craps$^a$ and Oleg Evnin$^b$}
\vskip 7mm
{\em $^a$ Theoretische Natuurkunde, Vrije Universiteit Brussel and\\
The International Solvay Institutes\\ Pleinlaan 2, B-1050 Brussels, Belgium}
\vskip 3mm
{\em $^b$ Institute of Theoretical Physics, Academia Sinica\\
Zh\=onggu\=anc\=un d\=ongl\`u 55, Beijing 100190, China}
\vskip 3mm
{\small\noindent  {\tt Ben.Craps@vub.ac.be, eoe@itp.ac.cn}}
\end{center}
\vfill

\begin{center}
{\bf ABSTRACT}\vspace{3mm}
\end{center}

Important open questions in cosmology require a better understanding of the Big Bang
singularity. In string and matrix theories, light-like analogues of cosmological singularities
(singular plane wave backgrounds) turn out to be particularly tractable.
We give a status report on the current understanding of such light-like Big Bang models,
presenting both solved and open problems.

\vfill

\end{titlepage}

\section{Introduction}

The last two decades have witnessed enormous progress in our understanding of the composition and evolution of the universe. One of the remaining challenges is to understand how the very early universe reached a nearly homogeneous, nearly flat state with a specific spectrum of density perturbations consistent with present observations.

The most popular explanation is that the very early universe underwent a period of inflation  \cite{I3_inflation}. If one assumes that inflation started and lasted long enough, it is able to explain the flatness and homogeneity of the universe. It also solves the monopole problem. The greatest success of inflation is that ``simple'' (single-field, slow-roll) inflationary models predict nearly scale-invariant, nearly Gaussian adiabatic density perturbations \cite{I3_perturbations}. These are the seeds of large scale structure and are visible as temperature anisotropies in the cosmic microwave background. One may wonder, though, how the universe emerged in a state that allowed inflation to start. In other words, how was a suitably fine-tuned initial state selected? In particular, in general relativity, inflationary solutions are past geodesically incomplete (under a certain assumption that excludes a contracting phase in the past) \cite{I3_pastincomplete}. The question should then be asked whether singularity resolution in a more fundamental theory puts constraints on which inflationary models are allowed.

Alternatives to inflation include the cyclic universe \cite{I3_cyclic}. The ekpyrotic mechanism (ultra-slow contraction) generates a spectrum of perturbations very similar to that of inflation, but in a contracting universe \cite{I3_ekpyroticperturbations}. In general relativity, the transition from a contracting to an expanding (spatially flat) universe requires going through a singularity \cite{I3_crunchbang}. At present, it is unclear whether such a transition is possible and whether perturbations would go through essentially unchanged. The answer will have to come from a theory beyond general relativity.

As we have argued, for the inflationary universe, and even more for alternative models, it is important to try and understand the big bang singularity. The work described in this review is motivated by several fundamental questions. Can we describe the big bang itself? How do space and time emerge from the big bang? Is it consistent to have a contracting universe before the big bang? Does the universe have a natural initial state, and if so, does it lead to inflation? String theory provides a short-distance modification of Einstein gravity, which is hoped to resolve space-time singularities. Existing formulations of string theory depend sensitively on the class of space-times one works with, and we will have to distinguish several classes of models. 

Perturbative string theory requires the background space-time to be specified from the onset. This background space-time has to satisfy supergravity equations of motion with an infinite number of corrections expanded in powers of $\alpha'$, the inverse of the string tension. In the high curvature regime, which necessarily accompanies singularities, all these $\alpha'$-corrections generically become equally important, and the background equations of motion generically become completely intractable. One exception is provided by orbifolds, obtained from manifolds by discrete identifications \cite{Dixon:1985jw}. Orbifolds contain new sectors of closed strings, namely ``twisted'' closed strings, which on the covering space connect a point and its image under a discrete identification. The rules of perturbative string theory on orbifolds are inherited from those on the covering space. If the discrete identifications have fixed points, orbifolds can be singular. It is known that static orbifold singularities are resolved in perturbative string theory, precisely thanks to the inclusion of twisted closed strings becoming light near the singular point. The hope 10 years ago was that time-dependent orbifolds (see \cite{Cornalba:2003kd, Durin:2005ix, Craps:2006yb, Berkooz:2007nm} for reviews) would lead to simple examples of cosmological singularities resolved within perturbative string theory. It turned out, however, that at least the simplest models were plagued by divergences related to large backreaction and invalidating string perturbation theory \cite{Liu:2002ft, Lawrence:2002aj, Horowitz:2002mw, Berkooz:2002je}.\footnote{See, however, \cite{Cornalba:2003kd} for a suggested resummation of divergences by working in the eikonal approximation, \cite{Cornalba:2002nv} for a proposed resolution of an orbifold singularity in terms of orientifolds, and \cite{McGreevy:2005ci} for a closely related model in which the singularity is replaced by a phase with a condensed winding tachyon within perturbative string theory.} It is worth noting that, in perturbative string theory, lightlike orbifold singularities are equally problematic as spacelike ones.

One can therefore turn to non-perturbative formulations of quantum gravity, such as the AdS/CFT-correspondence \cite{I3_AdS_CFT} or matrix \cite{Banks:1996vh} or matrix string \cite{Motl:1997th, Banks:1996my, Dijkgraaf:1997vv} theories. It should be said that those frameworks do not directly deal with the usual singularities discussed in cosmology. The AdS/CFT-correspondence requires the space-time to exhibit an AdS structure, at least asymptotically, while matrix and matrix string theories rely on the presence of a light-like isometry. None of these properties apply to the singularities naturally emerging in the context of classical cosmology: Friedmann, Kasner and Belinsky-Khalatnikov-Lifshitz space-times. Nevertheless, any progress on resolving light-like or space-like singularities, even in a not directly realistic model, would be very welcome, as it would point to mechanisms by which quantum gravity can in principle resolve cosmological singularities.

One clearly needs a compromise between the classes of space-times tractable within contemporary string theory and those relevant in cosmology. A number of directions of research have been proposed in this vein, and in this review we shall concentrate on one of them: analysis of light-like analogues of the usual singular cosmologies. For other models, we refer to the reviews \cite{Cornalba:2003kd, Durin:2005ix, Craps:2006yb, Berkooz:2007nm, McAllister:2007bg, Craps:2010bg}. Within the class of light-like singularities, we will focus on studies in perturbative string theory and matrix (string) theory. For studies of light-like singularities using the AdS/CFT correspondence, see the review \cite{Das:2007dw}.

The prototype metric of the kind we are going to consider can be written as
\be
ds^2=-2dx^+dx^-+\mu_{ij}(x^+)dx^idx^j.
\label{lightbang}
\ee
When $\mu$ is diagonal and depends on $x^+$ through power-law functions, this looks very much like a Kasner solution, except
that the dependences are on the light-cone time $x^+$ rather than on the usual time.
When $\mu$ is proportional to the unit matrix with a power-law dependence on $x^+$, the metric looks like a light-like version
of flat Friedmann cosmologies. Furthermore, (\ref{lightbang}) actually arises
when a Penrose limit is taken around a light-like geodesic hitting cosmological
singularities \cite{Blau:2004yi}.

The metric (\ref{lightbang}) is of the plane wave type, i.e., it describes a strong
plane-fronted gravitational wave (which is a non-linear generalization of the familiar
linearized gravitational waves in Minkowski space-time). Such plane waves are most
conveniently analyzed in the so-called Brinkmann coordinates, with their metric given by
\beq
ds^2=-2dx^+\,dx^-+K_{ij}(x^+)x^ix^j(dx^+)^2+(dx^i)^2,
\label{brink}
\eeq
where $K_{ij}(x^+)$ represent the profiles of different polarization components of the wave.
In pure gravity, $K_{ij}(x^+)$ is constrained by $K_{ii}=0$, giving the same number
of polarizations as in linearized theory (a traceless symmetric tensor in
$D-2$ dimensions, with $D$ being the number of dimensions of space-time).
If a dilaton is present, $K_{ii}$ does not vanish and is related to the dilaton,
which gives an additional independent polarization component. By a $u$-dependent
rescaling of $x^i$, (\ref{brink}) can be brought to the so-called Rosen form, in which the metric
only depends on $u$, making the planar nature of the wave front manifest. The Rosen form is
implied in (\ref{lightbang}). (It is prone to coordinate singularities, however, and often avoided.)

Plane wave space-times possess a number of remarkable properties in the context
of quantum gravity theories. In perturbative string theory, they are known to
satisfy background consistency conditions to all orders in $\alpha'$ \cite{Horowitz:1990sr}.
In other words, if supergravity equations of motion (zeroth order in $\alpha'$) are
satisfied, all the higher order corrections will vanish automatically. This property
comes from a special structure of the Riemann tensor in the plane wave background,
and it allows perturbative string analysis of even highly curved plane waves.
At the same time, the worldsheet theory of strings in plane wave backgrounds turns
out to be especially simple (and linear, when the light cone gauge is imposed). Similarly,
the light-like isometry needed for formulating matrix models is also present
in plane wave geometries, so a matrix theory description of quantum effects
in these space-times can be given.

It is most natural to start by studying (\ref{lightbang}) with $\mu_{ij}$ proportional
to the unit matrix (i.e., an isotropic space). Such space-times can be thought of
as light-like analogues of Friedmann cosmologies. By the equations of motion, such space-times
cannot be empty: one needs to add matter to compensate for the curvature of the plane
wave. A natural choice for this additional matter is the dilaton, a scalar always 
present in string theories and supergravities. If one starts with (\ref{lightbang}) and
takes $\mu_{ij}$ to be proportional to $\de_{ij}$ and depending on $x^+$ as a power law,
the corresponding Brinkmann form metric (\ref{brink}) can be written as
\beq
ds^2=-2dx^+\,dx^--\frac{k}{x^{+2}}(x^i)^2 (dx^+)^2+(dx^i)^2.
\label{lambda}
\eeq
In turn, the equations of motion determine the dilaton to be \cite{Papadopoulos:2002bg}
\be
\phi=\phi_0+cX^++\frac{kd}2\ln x^+
\label{dltn}
\ee
(where $d=D-2$ is the number of $i$-indices). Thus, for negative $k$, the string coupling $e^\phi$ blows up at the singularity ($x^+=0$), invalidating string perturbation theory. One would then expect a perturbative approach to be of little use for addressing the question of singularity transition in that case. It can still be applied, however, if $k>0$. (Singularities with positive $k$ arise as Penrose limits of power-law
Big Bang singularities, with the scale factor of the universe proportional to positive
powers of times; in the same way, $k<0$ corresponds to Big Rip singularities.)

It is 
often thought that quantum-gravitational effects should naturally resolve
singularities in some way. This would certainly be desirable, but is not so in our explicit examples. The singular plane wave (\ref{lambda}) enters various quantum
gravity constructions as the background. The singularity is then always there
at $x^+=0$, at least asymptotically, even though locally the geometry may be altered
(or even dissolved by non-geometrical states). Mathematically, the singularity
appears as explicit singular time dependence in the Hamiltonians of 
string and matrix theories in the plane wave (\ref{lambda}). One then has to understand
how to deal with such singular time dependences.

Free string propagation on (\ref{lambda}) was studied in \cite{Papadopoulos:2002bg}. In particular, it was suggested
in that publication that the question of propagation across the $1/(X^+)^2$
singularity in the metric can be addressed by employing analytic continuation
in the complex $X^+$-plane. Subsequently, in \cite{Craps:2008bv}, another principle was proposed, which we motivate next.\footnote{Recent work on closely related models appeared in \cite{Madhu:2009jh}.}

In string and matrix theories, it is necessary to satisfy stringent consistency
conditions in order to maintain finiteness and anomaly cancellation. In perturbative
string theories, this question is very well studied, and it is known that the space-time
background has to satisfy $\alpha'$-corrected supergravity equations of motion in
order for the theory to be well-defined (as already mentioned, the $\alpha'$ corrections
are absent if one is working with plane waves, hence satisfying plain supergravity equations of motion is sufficient). For matrix theories, similar restrictions
arise from considering $\kappa$-symmetry of the D-brane action \cite{Cederwall:1996ri}, though
the question does not appear to have been studied systematically. In any case, one would
expect that the handling of the singularity should be subject to rigid constraints
(given that even for smooth space-times one encounters rigid constraints).

There is one approach to handling plane wave singularities that automatically takes
benefit of what is known about smooth space-times and applies it to the singular limit.
Namely, one can consider (\ref{lambda}) as a limit of smooth metrics of the type (\ref{brink}), do the relevant computations, and take the singular limit at the end.
Then, for any resolved space-time (\ref{brink}) consistency of string theory is guaranteed
if (\ref{brink}) satisfies the supergravity equations of motion (without any further conditions). It is then natural to assume that the singular limit will likewise be a consistent string
theory, provided that this limit exists.

Even with these specifications, there are many ways to resolve (\ref{lambda}). One class of resolutions
appears to be very special however. The background (\ref{lambda})
possesses a scaling symmetry and does not depend on any dimensionful parameters.
It is natural to demand that this symmetry should be recovered when
the resolution is removed. This will happen if the resolved plane wave profile 
does not depend on any dimensionful parameters other than the resolution
parameter $\epsilon$. In this case, on dimensional grounds,
\be
K_{ij}(x^+,\epsilon)=-\de_{ij}\frac{1}{\epsilon^2}\Omega(x^+/\epsilon).
\label{scaleinvres}
\ee
The limit (\ref{lambda}) will be recovered if
\be
\Omega(\eta)\to \frac{k}{\eta^2}+O\left(\frac1{\eta^b}\right)
\label{asrefprof}
\ee
for large values of $\eta$, with some $b>2$.

In this review, we shall first concentrate on the perturbative string analysis of space-times (\ref{lambda}), and then discuss how this class of backgrounds can be treated in matrix and matrix string theories.

\section{Perturbative strings in singular plane wave backgrounds}

\subsection{The light cone Hamiltonian and WKB solutions}

String worldsheet fermions are free in plane wave backgrounds \cite{Russo:2002qj}. We shall therefore concentrate on the bosonic part of the string action, given by
\be
I=-\frac{1}{4\pi\alpha'}\int dt \int_0^{2\pi} d\sigma \sqrt{-g} \left(g^{ab} G_{\mu\nu} \partial_a X^\mu \partial_b X^\nu -\frac{1}{2} \alpha' R^{(2)}\Phi\right).
\label{bospartact}
\ee
We choose light-cone gauge $X^+=\alpha'p^+\tau$ and gauge-fix the worldsheet metric,
\be
\mathrm{det}(g_{ab}) = -1,\hspace{5mm}\partial_\sigma g_{\sigma\sigma}=0.
\ee
After some algebra and in units $\alpha'=1$, one obtains the light-cone Hamiltonian
as a sum of time-dependent harmonic oscillators:
\begin{align}
H&= \sum_{n=0}^\infty \sum_{i=1}^d H_{ni},\label{setham}\\
H_{0i}&=\frac{(P_{0i})^2}{2}+ \frac{1}{\epsilon^2}\Omega(t/\epsilon) \frac{(X_{0}^i)^2}{2},\label{HO0ham}\\
H_{ni}&=\frac{(P_{ni})^2+(\tilde{P}_{ni})^2}{2}+ \left(n^2 + \frac{1}{\epsilon^2}\Omega(t/\epsilon)\right) \frac{(X_{n}^i)^2+  (\tilde{X}_{n}^i)^2}{2},\label{HOham}
\end{align}
where $X_n$ are Fourier transforms in the $\sigma$-coordinate:
\be
X^i(t,\sigma)=X^i_0(t)+ \sqrt{2}\sum_{n>0} \left(\mathrm{cos}\left(n\sigma\right) X^i_{n}(t)+\mathrm{sin}\left(n\sigma\right) \tilde{X}^i_{n}(t)\right).
\ee

The Hamiltonian (\ref{setham}) is quadratic and the solution to the corresponding Schr\"odinger equation
can be found using WKB techniques, which are exact for quadratic Hamiltonians. The solution can be written as
\be
\phi_n^i(t;X)=\mathcal{A}_n(t_1,t)\,\mathrm{exp}\left(i S_{cl;n}[X_{1,n}^i,t_1|X_n^i,t]\right),
\label{WKB}
\ee
where $S_{cl;n}[X_{1,n}^i,t_1|X_n^i,t]$ is the ``classical action'' evaluated for the path going from $X_{1,n}^i$ at the time $t_1$ to $X_n^i$ at the time $t$,
\be
S_{cl}[X_{1,n}^i,t_1|X_n^i,t]=\int_{t_1}^{t} \mathrm{d}t' \left( \frac{(\dot{X_n^i})^2}{2}-\left(n^2+\frac{1}{\epsilon^2}\Omega\left(\frac{t'}{\epsilon}\right)\right)\frac{(X_n^i)^2}{2}\right).
\ee
If $\mathcal{A}_n(t_1,t)$ satisfies
\be
-2\frac{\partial}{\partial t}\mathcal{A}_n(t_1,t)=\mathcal{A}_n(t_1,t)\frac{\partial^2}{\partial (X_n^i)^2} S_{cl}[X_{1,n}^i,t_1|X_n^i,t],
\ee
then (\ref{WKB}) satisfies the original Schr\"odinger equation exactly.

Up to normalization, a basis of such solutions, labelled by the initial condition $X_n^i(t_1)=X_{1,n}^i$, is given by \cite{Evnin:2008ya}
\be
\phi(t;X_n^i) \sim \prod_{ni}\frac{1}{\sqrt{\mathcal{C}(t_1,t)}}\mathrm{exp}\left(-\frac{i}{2 \mathcal{C}}\sum_{i=1}^d \left[(X_{1,n}^i)^2\partial_{t_1}\mathcal{C}-(X_n^i)^2\partial_{t_2}\mathcal{C}+2 X_{1,n}^i X_n^i\right]\right),\label{basisofsols}
\ee
where $\mathcal{C}(t_1,t_2)$ (suppressing the index $n$) is a solution to the ``classical equation of motion'' for the time-dependent harmonic oscillator Hamiltonian (\ref{HOham}):
\be
\partial^2_{t_2}{\mathcal{C}}(t_1,t_2)+\left(n^2+\frac{1}{\epsilon^2}\,\Omega(t_2/\epsilon)\right)\mathcal{C}(t_1,t_2)=0,\label{ceomC}
\ee
with initial conditions specified as
\be
\mathcal{C}(t_1,t_2)|_{t_1=t_2}=0,\hspace{5mm}\partial_{t_2} \mathcal{C}(t_1,t_2)|_{t_1=t_2}=1.\label{initC}
\ee
We shall refer to $\mathcal{C}(t_1,t_2)$ as ``compression factor''.
To derive the singular limit of the wavefunction (\ref{basisofsols}) it is sufficient to study the singular limit of (\ref{ceomC}-\ref{initC}).


\subsection{The singular limit for the center-of-mass motion}
\label{0mode}

For the $n=0$ mode, we obtain as the ``classical equation of motion''
\be
\ddot{X}+\frac{1}{\epsilon^2}\Omega(t/\epsilon)X=0.
\ee
We need to study the $\epsilon\rightarrow 0$ limit of the solution that obeys the initial conditions
\be
X(t_1)=0,\hspace{10mm}\dot{X}(t_1)=1,\hspace{10mm}t_1<0. 
\label{initcondX}
\ee

The singular limit of solutions to this equation has been analyzed in \cite{Evnin:2008ya}. Performing a scale transformation $Y(\eta)=X(\eta\epsilon)$, with $\eta=t/\epsilon$, removes the $\epsilon$-dependence from the equation, leaving
\be
\frac{\partial^2}{\partial \eta^2}{Y}+\Omega(\eta)Y=0.
\label{etaEq}
\ee
This scale transformation is possible because our initial singular metric was scale-invariant and we have resolved it as in (\ref{scaleinvres}) without
introducing any dimensionful parameters besides $\epsilon$. The existence of a singular limit is then translated \cite{Evnin:2008ya} into constraints on the asymptotic behavior of solutions to (\ref{etaEq}). These ``boundary conditions at infinity'' are strongly reminiscent of a Sturm-Liouville problem, and it is natural 
that a discrete spectrum for the overall normalization of $\Omega$ is singled out by imposing the existence
of a singular limit.

For the specific asymptotics of our resolved profile (\ref{asrefprof}), it can be shown \cite{Evnin:2008ya}
that, in the infinite past and infinite future, the solutions approach a linear combination of two powers (denoted below $a$ and $1-a$, with $a$ being a function
of $k$, cf.\ (\ref{scaleinvres}-\ref{asrefprof})). This power law behavior simply corresponds to the regime when
the second term on the right hand side of (\ref{asrefprof}) can be neglected compared to the first. It is then convenient to form two bases of solutions, one asymptotically approaching the two powers (dominant and subdominant) at $\eta\rightarrow -\infty$,
\be
Y_{1-}(\eta)=|\eta|^{a_-} + o(|\eta|^{a_-}),\hspace{10mm}Y_{2-}(\eta)=|\eta|^{1-a_-} + o(|\eta|^{1-a_-}),
\label{Y-}
\ee
and another behaving similarly at $\eta\rightarrow +\infty$
\be
Y_{1+}(\eta)=|\eta|^{a_+} + o(|\eta|^{a_+}),\hspace{10mm}Y_{2+}(\eta)=|\eta|^{1-a_+} + o(|\eta|^{1-a_+}),
\label{Y+}
\ee
where $a_\pm$ is given by
\be
a_\pm=\frac12+\sqrt{\frac14- k_\pm}.
\ee
(We are temporarily assuming that $k$ can take two different values $k_\pm$ for
the positive and negative time asymptotics, a possibility that will be discarded shortly.) The two bases are related by a linear transformation:
\be
\begin{bmatrix}Y_{1-}(\eta)\\Y_{2-}(\eta)\end{bmatrix}=Q\begin{bmatrix}Y_{1+}(\eta)\\Y_{2+}(\eta)\end{bmatrix},
\label{Q}
\ee
where $Q$ is a $2\times 2$ matrix whose determinant is constrained by Wronskian conservation as
\be
W[Y_{1-},Y_{2-}]=W[Y_{1+},Y_{2+}]\det Q .
\ee

The singular limit has been rigorously considered in \cite{Evnin:2008ya},
but the results can be understood heuristically from the following argument \cite{Craps:2008bv}.
Imagine one is trying to construct a solution $\tilde Y$ to (\ref{etaEq}) satisfying
some ($\epsilon$-independent) initial conditions at $\eta_1=t_1/\epsilon<0$. This solution can be
expressed in terms of $Y_{1-}$ and $Y_{2-}$ (a complete basis) as
\be
\tilde Y=C_1 Y_{1-}+C_2 Y_{2-}.
\ee
Since the initial conditions are specified at $\eta_1=t_1/\epsilon$, the asymptotic
expansions (\ref{Y-}) are valid. There needs to be a non-trivial contribution
from both $Y_{1-}$ and $Y_{2-}$ in the above formula in order to satisfy general
initial conditions. Hence, the two terms on the right hand side should be of order 1.
Therefore, we should have
\be
C_1=O(\epsilon^{a_-}),\qquad C_2=O(\epsilon^{1-a_-}).
\ee
If we now apply (\ref{Q}) and (\ref{Y+}) to evaluate $\tilde Y$ at a large positive $\eta=t_2/\epsilon$, the powers of $\epsilon$ in $C_1$ and $C_2$ will combine
with the powers of $\epsilon$ originating from $Y_{1+}$ and $Y_{2+}$ and yield
\be
\begin{array}{l}
\dsty\tilde Y(t_2/\epsilon)=Q_{11}t_2^{a_+}O(\epsilon^{a_--a_+})+Q_{12}t_2^{1-a_+}O(\epsilon^{a_-+a_+-1})\vspace{2mm}\\
\dsty\hspace{3cm}+Q_{21}t_2^{a_+}O(\epsilon^{1-a_--a_+})+Q_{22}t_2^{1-a_+}O(\epsilon^{a_+-a_-}).
\end{array}
\ee
Since $a_+$ and $a_-$ are greater than $1/2$, this expression can only have an $\epsilon\to 0$ limit if $a_+=a_-$ (i.e., $k_+=k_-$) and $Q_{21}=0$.
The latter condition implies that the overall normalization of the plane wave profile $\Omega(\eta)$ will generically lie in a discrete spectrum, dependent on the specific way the singularity is resolved, i.e., the detailed shape of $\Omega(\eta)$. A particular exactly solvable example for this discrete spectrum (there called ``light-like reflector plane'') has been given in \cite{Craps:2008cj}. With $Q_{21}=0$ and $\det Q=-1$, the matrix $Q$ can be written as
\be
Q=\begin{bmatrix}q&\tilde q\\0&-1/q\end{bmatrix},
\label{asbas}
\ee
with $q$ being a real nonzero number ($\tilde q$ does not affect the singular limit). For flat space-time we have $q=1$ and for the ``light-like reflector plane'' of \cite{Craps:2008cj} we have $q=-1$. In the singular limit, a basis of solutions is given by
\begin{align}
&Y_{1}(t)=(-t)^a,\hspace{5mm}Y_{2}(t)=(-t)^{1-a},\hspace{5mm}t<0,\nonumber\\
&Y_{1}(t)=q\,t^a,\hspace{5mm}Y_{2}(t)=-\frac{1}{q}t^{1-a},\hspace{5mm}t>0.\label{basissolzeromode}
\end{align}

\subsection{The singular limit for excited string modes}

Following our general discussion of free strings in plane wave backgrounds,
the evolution of excited string modes is described by time-dependent harmonic oscillator equations
\be
\frac{\partial^2}{\partial t^2} X(t) + \left(n^2+\frac{1}{\epsilon^2}\Omega(t/\epsilon)\right) X(t)=0.\label{de}
\ee
Solutions for the wavefunctions of the excited string modes can be expressed in
terms of a particular solution to this equation $\mathcal{C}(t_1,t_2)$ defined by (\ref{ceomC}-\ref{initC}). Hence, to analyze the singular ($\epsilon\to 0$) limit of the excited modes dynamics, it should suffice to analyze the singular limit of $\mathcal{C}(t_1,t_2)$. Because $n^2$ is finite, it is natural to expect that it does not affect the existence of the singular limit (governed by the singularity emerging from $\Omega(t/\epsilon)$). It can be proved that it is indeed the case for positive $k$ \cite{Craps:2008bv}.

The general strategy here is to analyze (\ref{de}) separately in the near-singular region ($t$ close to 0) and the region where the $\eps\to 0$ limit is regular.
Away from $t=0$, up to corrections vanishing as $\eps$ is taken to 0, (\ref{de}) can be
approximated by
\beq
\ddot{X}(t) + \left(n^2+ \frac{k}{t^2}\right) X(t)=0,
\label{de_app1}
\eeq 
which is related to Bessel's equation. Around $t=0$, one should expect that
$n^2$ can be neglected, which leaves the equation for the zero-mode (already
analyzed in the previous section):
\beq
\ddot{X}(t) + \frac{1}{\epsilon^2} \Omega(t/\epsilon) X(t)=0.
\label{de_app2}
\eeq
More specifically, the separation into near-singular and regular regions should
be organized as follows:

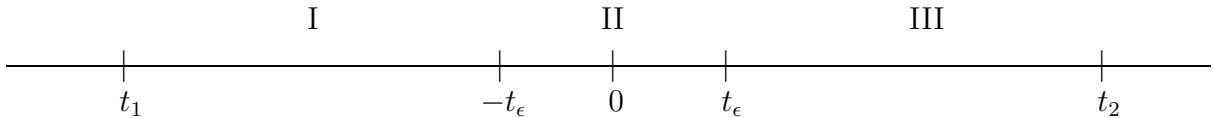
\setlength{\unitlength}{1mm}
\noindent\begin{picture}(160,20)
\put(0,10){\line(1,0){160}}
\put(40,15){\text{I}}
\put(79,15){\text{II}}
\put(120,15){\text{III}}
\put(15,9){\text{$|$}}
\put(65,9){\text{$|$}}
\put(95,9){\text{$|$}}
\put(145,9){\text{$|$}}
\put(15,4){\text{$t_1$}}
\put(63,4){\text{$-t_\epsilon$}}
\put(95,4){\text{$t_\epsilon$}}
\put(145,4){\text{$t_2$}}
\put(80,9){\text{$|$}}
\put(80,4){\text{$0$}}
\end{picture}

\noindent We use $t_\epsilon$ to indicate a time that will approach zero in the singular limit as 
\be
t_\epsilon=\epsilon^{1-c}\tilde t^c,
\label{teps}
\ee
with $\tilde t$ staying finite in relation to the ``moments of observation'' $t_1$ and $t_2$. The number $c$ (between 0 and 1) should be chosen later as needed for our proof.

One then can show that (for positive $k$) there is indeed a choice of $c$ that
makes deviations from the approximate equations (\ref{de_app1}-\ref{de_app2})
small in the appropriate regions. One can then construct the $\eps\to 0$ limit
of solutions to (\ref{de}) by taking approximate solutions satisfying (\ref{de_app1}) and (\ref{de_app2}), splicing them together and taking the $\eps\to 0$ limit in the end.
Thus, one obtains exact expressions for the $\eps\to 0$ limit of solutions to (\ref{de}),
even though analytic solutions to (\ref{de}) at finite $\eps$ cannot be given.

We refer the reader to the original article \cite{Craps:2008bv} for detailed proofs,
and here simply state the result:
For $k>0$, the singular limit of the excited mode evolution exists whenever it
exists for the center-of-mass motion, and it is described by the following
matching conditions:
\begin{align}
&Y_{1}(t)=\sqrt{-t}J_{a-1/2}(-n t),\hspace{5mm}Y_{2}(t)=\sqrt{-t}J_{1/2-a}(-n t),\hspace{5mm}t<0,\nonumber\\
&Y_{1}(t)=q\sqrt{t}J_{a-1/2}(n t),\hspace{5mm}Y_{2}(t)=-\frac{\sqrt{t}}{q}J_{1/2-a}(n t),\hspace{5mm}t>0\label{basissol},
\end{align}
where $J_\nu$ are Bessel functions.

Note that there is a slight subtlety in the sense that convergence to the $\eps\to 0$ 
limit is not uniform with respect to $n$ (we have kept $n$ fixed in our considerations).
However, this does not affect the result as long as one is only interested in the limiting expressions at $\eps=0$. More discussion is given in the original publication \cite{Craps:2008bv}.


\subsection{The singular limit for the entire string}
\label{total}

As we have seen in the previous section, for $k >0$, 
consistent propagation of the string
center-of-mass across the singularity guarantees that all excited string modes
also propagate in a consistent fashion. This is not sufficient, however,
to define a consistent evolution for the whole string, since even small
excitations of higher string modes can sum up to yield an infinite total
energy \cite{Horowitz:1990sr}. As we shall see below, the condition of finite total string energy
(after the singularity crossing) turns out to be very restrictive.

The total string excitation energy can be conveniently expressed in terms
of the Bogoliubov coefficients for the higher string modes which can be extracted
from (\ref{basissol}) as
\begin{align}
\alpha_n&=-\frac{1+q^2}{2 q \,\mathrm{sin}(\alpha\pi)},\\
\beta_n&= i\frac{\mathrm{exp}(-i\pi \alpha)+q^2\mathrm{exp}(i\pi \alpha)}{2 q \,\mathrm{sin}(\alpha\pi)},
\end{align}
and they turn out to be independent of $n$. Here, $\alpha=\sqrt{1-4k}/2$. The total mass of the string after crossing the singularity is given by \cite{Horowitz:1990sr}
\be
M=\sum_n n |\beta_n|^2.
\label{Msum}
\ee
Since the $\beta_n$ are $n$-independent, $M$ can only be finite if $\beta_n=0$ for all $n$. For $k>0$, this cannot be achieved, since $0<\alpha<1/2$ and $q$ is real.
(For $k=0$, which is the case of the ``lightlike reflector plane'' of \cite{Craps:2008cj},
all $\beta_n$ will vanish if $q^2=1$, which is satisfied automatically for any reflection-symmetric $\Omega$.) The implication is then that, if the singularity is resolved without introducing any new dimensionful scales, the singularity transition cannot be defined. (The alternative is having dimensionful parameters buried strictly
at the singular locus, and an explanation of the possible physical origin of such parameters would be in order.)

As mentioned before, in the case $k<0$ the string coupling blows up near the singularity, so that string perturbation theory is certainly not valid. Our considerations can be seen as a motivation to study these backgrounds in the context of non-perturbative formulations of string theory, to which we now turn.

\section{Matrix big bang models}

\subsection{Time-dependent matrix and matrix string theories}

Matrix theory \cite{Banks:1996vh} is a non-perturbative formulation of M-theory in 11-dimensional Minkow\-ski space-time. One way to derive it is by Discrete Light-Cone Quantization (DLCQ) \cite{Susskind:1997cw}. A sector with $N$ units of lightcone momentum is described by a quantum mechanics of $N\times N$ matrices, namely the dimensional reduction of (9+1)-dimensional $SU(N)$ super-Yang-Mills theory to 0+1 dimensions. In matrix theory, (lightcone) time is built in, but space is an emergent concept, arising from a ``moduli space'' of flat directions. For this ``moduli space'', and therefore space-time, to emerge, supersymmetry plays an essential role: without supersymmetry, quantum corrections would lift the flat directions. Compactifying M-theory on a circle leads to type IIA string theory. Matrix string theory \cite{Motl:1997th, Banks:1996my, Dijkgraaf:1997vv} is a non-perturbative formulation of type IIA string theory in 10d Minkow\-ski space-time. The aspects of matrix and matrix string theory needed for our purposes are reviewed in detail in \cite{Craps:2006yb}.

The construction of matrix theories essentially relies on compactifying a light-like
direction (which is a pre-requisite for discrete light-cone quantization). Due to
discreteness, positivity and conservation of the light-cone momenta in such a compactified
space-time, Hilbert spaces of quantum theories living in this space-time split into
independent sectors (labelled by the value of the light-cone momentum), each of which
correponds to a quantum system with a {\it finite} number of degrees of freedom.
For matrix theories, this quantum system is given explicitly by finite $N$ matrix
Lagrangians. This is a remarkable simplification (which automatically renders the
theory UV-finite, since there are no divergences in quantum mechanics). The drawback
is equally grave, however, as reconstructing quantities in an infinite space-time
from the compactified version is highly non-trivial (it is believed to be possible,
however, since for a sufficiently large compactification radius, an arbitrarily large
laboratory can fit into the space-time with a light-like compactification).

Compactifying a light-like direction is only possible in space-time backgrounds
with a light-like isometry. This obviously excludes the case of ordinary cosmological
space-times, as well as almost all familiar non-trivial solutions to the classical
gravitational equations of motion. Strong plane waves of the type (\ref{lightbang}, \ref{brink}) are an exception though, which highlights once again their special status
in quantum gravity and makes them an interesting laboratory for applying matrix
theory methods to study time dependence and strong gravitational effects.

One must keep in mind that reconstructing quantities in a decompactified space-time
from their DLCQ analogues (explicitly given by the finite $N$ matrix theories)
may become more subtle in a time-dependent setting.\footnote{We thank David Kutasov
for drawing our attention to this issue.} The point is that, at best, DLCQ can
provide information on quantities in a decompactified space that are measurable in a finite-size
laboratory. This is not likely to pose a problem for the case of mild time dependences.
But for the opposite extreme, namely space-times including singularities (which we shall comment
on in the next section), linear dimensions of physical systems in an infinite space
may blow up indefinitely. In that case, finite box dynamics (and hence DLCQ) will
not be adequate to describe the evolution in a decompactified space. Whether or
not these subtleties do arise has to be decided on the basis of careful dynamical
considerations, which have not been carried out yet.

We shall now give a summary of a few different versions of matrix and matrix string theories
in plane wave backgrounds.
For the original time-dependent Matrix Big Bang
matrix string theory of \cite{Craps:2005wd}, the 10-dimensional geometry is asymptotic
to the linear dilaton configuration:\footnote{The original set-up of \cite{Craps:2005wd} has been
later extended in various directions \cite{Li:2005sz,Li:2005ti,Das:2005vd,Chen:2005mga,Robbins:2005ua,Das:2006dr,Martinec:2006ak,Chen:2006rm,Ishino:2006nx,Bedford:2007dd,Blau:2008bp}; in particular, a systematic generalization of the analysis to more general singular homogeneous plane-wave space-time backgrounds has appeared in
\cite{Blau:2008bp}.}
\be
\begin{aligned}
ds_{st}^2 &= -2 dy^+ dy^- + (dy^i)^2,\\
\phi &= -Qy^+.
\end{aligned}
\label{mbb}
\ee
To construct the matrix string theory for the background (\ref{mbb}), one first lifts the background (\ref{mbb}) to 11 dimensions via the usual conjecture of type IIA/M-theory correspondence. The resulting 11-dimensional space-time is
\be\label{11met}
ds^2 = e^{ 2 Q y^+/3 } \left(-2 dy^+ dy^- + (dy^i)^2\right) + e^{-4 Q y^+/3} (dy)^2,
\ee
where $y$ is a coordinate along the M-theory circle. This is followed by the DLCQ compactification of the light-like $v$-coordinate, interpreted as the M-theory circle of an ``auxiliary'' type IIA string theory. A T-duality \cite{Taylor:1996ik} then relates the resulting theory of D0-branes on a compact dimension, i.e., a BFSS-like matrix theory with a compactified dimension, to a more manageable theory of wrapped D1-branes. This procedure has been carried out (in a slightly different but equivalent way) in \cite{Craps:2005wd} and has been reviewed in \cite{Craps:2006yb} and \cite{Blau:2008bp}. The resulting matrix string action is 
\be
\label{sym} S = \frac{1}{2\pi \ell_s^2}
\int {\tr}\left( \frac12(D_\mu X^i)^2 + \theta^T
{D\!\!\!\!\slash{}} \,\theta + \frac1{4g_{YM}^2} F_{\mu\nu}^2 - g_{YM}^2[X^i,X^j]^2 + g_{YM} \theta^T\gamma_i
[X^i,\theta]\right),
\ee
with the Yang-Mills coupling $g_{YM}$ related to the worldsheet values of the dilaton:
\be
g_{YM}=\frac{\ex{-\phi(y^+(\tau))}}{2\pi l_sg_s}=\frac{e^{Q\tau}}{2\pi l_sg_s}.
\ee

A generalization of this set-up has been proposed \cite{Blau:2008bp}. One can start with a 
10-dimensional power-law plane wave:
\beq
\begin{aligned}
\label{bn1}
ds_{st}^2 &= -2 dy^+ dy^- + g_{ij}(y^+) dy^i dy^j\equiv-2 dy^+ dy^- + \sum_{i} (y^+)^{2m_i} (dy^i)^2\\
&= -2dz^+dz^- + \sum_a \frac{m_a(m_a-1)}{(z^+)^2}(z^a)^2 (dz^+)^2 
+ \sum_a (dz^a)^2,\\
\ex{2\phi} &= (y^+)^{3b/(b+1)}=(z^+)^{3b/(b+1)}.
\end{aligned}
\eeq
Here, the first and the second line represent the Rosen and the Brinkmann form of the same plane wave, respectively. In order for the supergravity equations of motion to be satisfied, one needs to impose \cite{Blau:2008bp}
\be
\label{ede2}
\sum_i m_i(m_i-1) = -\frac{3b}{b+1}.
\ee
The original background of \cite{Craps:2005wd} can be seen as a $b\to-1$ limit of the above space-time \cite{Blau:2008bp}.
The 11-dimensional space-time corresponding to (\ref{bn1}) is
\be
\label{rcpl2}
\begin{aligned}
ds_{11}^2 &= -2dudv + \sum_i u^{2n_i}(dy^i)^2 + u^{2b}(dy)^2\\
&= - 2du dw + \sum_a \frac{n_a(n_a-1)}{u^2}(x^a)^2 (du)^2 +
\frac{b(b-1)}{u^2}x^2 (du)^2 + \sum_a (dx^a)^2 + (dx)^2, \\
\end{aligned}
\ee
with $n_i$ related to $m_i$ by $2m_i = (2n_i+b)/(b+1)$. 
The usual formulation leads to a matrix string action, whose bosonic part is given, in the Rosen coordinates of (\ref{bn1}), by
\begin{multline}
\label{src}
S_{RC} = \int d\tau d\sigma\Tr\left( -\frac{1}{4}g^{-2}_{YM}
\eta^{\alpha\gamma}\eta^{\beta\delta}F_{\alpha\beta}F_{\gamma\delta}
-\frac{1}{2}\eta^{\alpha\beta}g_{ij}(\tau)D_{\alpha}X^i D_{\beta}X^j
\right. \\ \left. 
+ \frac{1}{4}g^2_{YM} g_{ik}(\tau)g_{jl}(\tau)[X^i,X^j][X^k,X^l] \right)
.
\end{multline}
with the transverse metric $g_{ij}$ given by the first line of (\ref{bn1}) and the Yang-Mills coupling by
\be
g_{YM}=\frac{\ex{-\phi(y^+(\tau))}}{2\pi l_sg_s}=\frac{\tau^{-3b/2(b+1)}}{2\pi l_sg_s}.
\ee
One can further transform this action to the Brinkmann coordinates of the 
original plane wave, given by the second line of (\ref{bn1}), to obtain \cite{Blau:2008bp}:
\begin{multline}
\label{sbc}
S_{BC} = \int d\tau d\sigma\Tr\left( -\frac{1}{4}g^{-2}_{YM}
F_{\tau\sigma}^2
-\frac{1}{2}\left(D_{\tau}Z^a D_{\tau}Z^a-D_{\sigma}Z^a D_{\sigma}Z^a\right)\right. \\
\left. + \frac{1}{4}g^2_{YM} [Z^a,Z^b][Z^a,Z^b]  +
\frac{1}{2}A_{ab}(\tau)Z^aZ^b \right),
\end{multline}
where $A_{ab}=\mbox{diag}\{m_a(m_a-1)\}/\tau^2$. The latter form of the action only differs from a SYM gauge theory with a time-dependent coupling by the term involving
$A_{ab}$.

In \cite{Li:2005sz}, 11-dimensional (quantum-mechanical)
matrix theories were introduced as simpler analogues of the 
matrix string theories of \cite{Craps:2005wd}. The relevant 11-dimensional
(M-theory) background has the form
\beq
ds^2=e^{2\alpha x^+}\left(-2dx^+dx^-+(dx^i)^2\right)+e^{2\beta x^+}(dx^{11})^2,
\eeq
or, in terms of the light-like geodesic affine parameter $\tau=e^{2\alpha x^+}/2\alpha$,
\beq
ds^2=-2d\tau dx^-+2\alpha\tau(dx^i)^2+(2\alpha\tau)^{\beta/\alpha}(dx^{11})^2.
\label{bckgr}
\eeq
This metric satisfies the 11-dimensional supergravity equations of motion
if the constants $\alpha$ and $\beta$ are related as $\beta=-2\alpha$,
or $\beta=4\alpha$. The fact that these relations need to be imposed
will not be relevant for what follows (it is essential,
however, for the general consistency of the corresponding matrix theories).
Since translations in $x^-$ form an isometry of the above background,
the usual DLCQ argument (proposed in \cite{Seiberg:1997ad} and adapted to the
time-dependent case in \cite{Craps:2005wd}) can be applied. The result \cite{Li:2005sz}
is a matrix theory that can be expected to describe non-perturbative
quantum gravity in space-times asymptotic to (\ref{bckgr}). The bosonic
and fermionic parts of the matrix theory action, respectively,
have the following form:
\beq
\ber{l}
\dsty S_B=\int d\tau\text{Tr}\left\{\frac{\alpha\tau}{R}(D_\tau X^i)^2+\frac{(2\alpha\tau)^{\beta/\alpha}}{2R}
(D_\tau X^{11})^2- \frac{R}{4}(2\alpha\tau)^2[X^i,X^j]^2\right.\vspace{2mm}\\
\dsty\hspace{2cm}\left.-\frac{R}{2}(2\alpha\tau)^{1+\beta/\alpha} [X^i,X^{11}]^2\right\},\vspace{3mm}\\
S_F=\int d\tau\left\{i\theta^TD_\tau\theta -R\sqrt{2\alpha\tau}\theta^T\gamma_i
[X^i, \theta ]-R(2\alpha\tau)^{\beta/2\alpha}\theta^T\gamma_{11} [X^{11}, \theta ]\right\}.
\eer
\label{mlact}
\eeq

\subsection{Singularity transition}

The models we have just formulated are aimed to describe quantum-gravitational effects
in singular plane waves (\ref{rcpl2}). In Rosen coordinates (first line of (\ref{rcpl2})),
these space-times can be seen as a light-like analogue of Friedmann or Kasner cosmologies.
It is then interesting to investigate what the theory has to say about the question
of singularity transition.

The general structure that emerges from our considerations is matrix (string) Hamiltonians
with singular time-dependences. This can be explicitly seen in (\ref{sbc}) as $A_{ab}$
is singular at $\tau=0$. Likewise, the factors of $\tau$ in front of the time derivatives
in (\ref{mlact}) will become inverted when the velocities are replaced by momenta, and
the corresponding Hamiltonian will contain an isolated singularity in its time dependence
at $\tau=0$. These isolated singularities in the explicit time dependence of the matrix
(string) Hamiltonians are directly inherited from the singularity in the background
space-time. (Such singularities are also present in Hamiltonians (\ref{HO0ham}-\ref{HOham})
in the perturbative string theory setting of the previous section, if one looks
at the $\eps\to 0$ singular limit directly without performing the singularity resolution.)

One might have thought that quantum gravity should resolve singularities in some way
and give a dynamical prescription for singularity transitions. Unfortunately, it is
not so in our present setting. The underlying space-time singularities appear as 
explicit singularities in time dependences of matrix (string) Hamiltonians, and
the evolution is not well-defined without additional prescriptions.

Some general properties of Hamiltonians with singular time-dependences have been
studied in \cite{sing, Craps:2008cj}. An important thing to understand is that there is a tremendous
ambiguity associated with defining the singularity transition in this context.
Indeed, if one only assumes that the Schr\"odinger equation for the Hamiltonian
with a singular time dependence is satisfied away from singularity, and finds a way
to define the singularity transition, one can immediately create another singularity
transition prescription with the same properties. Namely, one can change the wave vectors
by an arbitrary unitary transformation the moment they pass the singularity. The Schr\"odinger equation will still be satisfied away from the singularity, as it was before.
One is, therefore, in a need of a physical (or at least heuristic) motivation for
choosing a particular prescription for singularity transition.

We believe that a very natural class of singularity transitions arises from
considering geometrical resolutions of the singular plane wave (\ref{rcpl2}).
Due to the functional arbitrariness of the plane wave profile,
it is very easy to replace the singular profile in (\ref{rcpl2}) by a resolved one
in a way similar to (\ref{scaleinvres}-\ref{asrefprof}). Maintaining the background
space-time as a solution to the equations of motion is also essential from the standpoint of background consistency. Enforcing supergravity equations for the background appears to be related to the $\kappa$-symmetry of the D-brane action \cite{Cederwall:1996ri} necessary for the
standard formulation of matrix theories. Unlike the case of perturbative strings,
the action is non-linear, so the singular limit may be much more difficult to investigate,
and it remains an important open problem for the future.

(Note the similarities of this setting to the AdS light-like cosmologies reviewed in \cite{Das:2007dw}. There, one ends up with a gauge theory featuring singular light-like time dependences. The only difference in our case is that the singular dependences are
on ordinary time. Similar non-linearities characteristic of gauge theories are present in both cases. In \cite{Das:2007dw}, a special class of plane waves is chosen for which the analysis simplifies due to conformal flatness. General p-brane-plane-wave solutions presented in \cite{pbrane} suggest a natural generalization of AdS light-like cosmologies
featuring arbitrary plane wave profiles. In that setting, the problem of singularity transition is even more similar to what one encounters in Matrix Big Bang models.)

\subsection{Near-classical late time dynamics}

Even though, in the context of time-dependent matrix and matrix string theories, novel physics is expected to emerge in the high-curvature regions of space time, it is also
important to understand how the near-classical space-time emerges away from the
singularity when the curvature becomes small. The issue may be technically 
somewhat involved, since geometrical notions appear rather indirectly in the matrix formalism. Heuristic results appeared in \cite{Craps:2005wd, Craps:2006xq}; a systematic analysis has recently been carried out in \cite{Craps:2010cn}, the results of which we now review.\footnote{Another attempt to study late time (low background
curvature) dynamics of the time-dependent matrix theories was undertaken
in \cite{O'Loughlin:2010uh}.} 

The problems of late time (near-classical) dynamics in matrix models can be understood by inspecting the action (\ref{mlact}). The space-time backgrounds implicit in
(\ref{mlact}) are supposed to feature a light-like singularity at $\tau=0$
and become progressively more classical at large $\tau$. Yet, the explicit
time dependences in (\ref{mlact}) superficially become more steep, if anything,
at large $\tau$. Additionally,
there are no supersymmetries explicit in (\ref{mlact}). Since supersymmetries
are crucial for the free propagation of well-separated gravitons (and hence, a
robust geometrical interpretation) in the flat space matrix theory, one should
attempt to find an analogue of supersymmetry in (\ref{mlact}) that would enforce
a similar type of dynamics.

To address these important issues, we first note that the metric (\ref{bckgr})
describes a plane wave and, with the coordinate transformation
\beq
u=\tau,\quad z^i=\sqrt{2\alpha\tau}x^i,\quad z^{11}=(2\alpha\tau)^{\beta/2\alpha}x^{11},\quad v=x^-+\frac{\alpha(z^i)^2+\beta(z^{11})^2}{4\alpha\tau},
\label{coord}
\eeq
it can be brought to the Brinkmann form
\beq
ds^2=-2du\,dv-\frac{\alpha^2(z^i)^2-(\beta^2-2\alpha\beta)(z^{11})^2}{(2\alpha u)^2}du^2+(dz^i)^2+(dz^{11})^2.
\label{brinkmn}
\eeq
This parametrization forces the metric to manifestly approach Minkow\-ski space-time
for large values of the light-cone time, which strongly suggests that the large
time dynamics of the corresponding matrix theory will likewise approach the flat
space matrix theory, if treated in appropriate variables. (As we shall see below, the convergence towards this limit is somewhat subtle, but the na\"\i ve
expectation will prove well-grounded.)

The matrix theory corresponding to (\ref{brinkmn}) is given by
\beq
\ber{l}
\dsty S_B=\int d\tau\text{Tr}\left\{\frac{1}{2R}\left[(D_\tau Z^i)^2+
(D_\tau Z^{11})^2\right]- \frac{R}{4}\left[[Z^i,Z^j]^2+2 [Z^i,Z^{11}]^2\right]\right.\vspace{2mm}\\
\dsty\hspace{2cm}\left.-\frac{\alpha^2(Z^i)^2-(\beta^2-2\alpha\beta)(Z^{11})^2}{(2\alpha \tau)^2}\right\},\vspace{3mm}\\
S_F=\int d\tau\left\{i\theta^TD_\tau\theta -R\theta^T\gamma_i
[Z^i, \theta ]-R\theta^T\gamma_{11} [Z^{11}, \theta ]\right\}.
\eer
\label{mlactbrnk}
\eeq

The action (\ref{mlactbrnk}) only differs from the flat space matrix theory
by a term decaying as $1/\tau^2$, thus one may expect that the late time
dynamics will be approximated by the flat space matrix theory and admit
the usual space-time interpretation. However, the decay is quite slow
and one might be worried about whether it is sufficient to ensure convergence.

To illustrate these worries, one may look at the straightforward example
of a harmonic oscillator whose frequency depends on time as $1/t^2$:
\beq
\ddot x + \frac{k}{t^2}x=0.
\label{ho}
\eeq
The two independent solutions to this equation can be given as $t^a$ and $t^{1-a}$,
where $a$ is a $k$-dependent number. These two solutions are obviously quite
different from a free particle trajectory, even though the equation of motion
approaches that of a free particle at late times. The reason for this discrepancy
is the slow rate of decay of the second term in (\ref{ho}).

However, in a physical setting, one is only able to perform {\it finite time} experiments.
That is, one has to specify the initial values $x(t_0)=x_0$, $\dot x(t_0)=v_0$
and examine the corresponding solution between $t_0$ and $t_0+T$. The solution is given by
\beq
x(t) = \frac{x_0(1-a)-v_0t_0}{1-2a}\left(\frac{t}{t_0}\right)^a + \frac{v_0t_0-x_0a}{1-2a}\left(\frac{t}{t_0}\right)^{1-a}.
\eeq
One can then see that $x(t_0+T)=x_0+v_0T+O(T/t_0)$, i.e., it is approximated
by a free motion arbitrarily well if the experiment starts sufficiently late.

It may be legitimately expected that the {\it finite time} behavior of the full time-dependent matrix theory given by (\ref{mlactbrnk}) will be approximated arbitrarily
well by the flat space matrix theory at late times, just as in the above harmonic
oscillator example. We shall now prove it by constructing an elementary bound
on dynamical deviations due to a small time-dependent term in the Schr\"odinger
equation.

We start with the following Schr\"odinger equation:
\beq
i\frac{d}{dt}\left|\Phi\right>=\left(H_0+f(t)H_1\right)\left|\Phi\right>,
\eeq
where $H_0$ and $H_1$ are time-independent, and rewrite it in the interaction picture (with respect to $H_0$):
\beq
\left|\Phi\right>=e^{-iH_0(t-t_0)}\left|\xi\right>,\qquad
i\frac{d}{dt}\left|\xi\right>=f(t)e^{iH_0(t-t_0)}H_1e^{-iH_0(t-t_0)}\left|\xi\right>.
\eeq
We then proceed to consider
\beq
\ber{l}
\dsty\frac{d}{dt}\Big|\left|\xi(t)\right>-\left|\xi(t_0)\right>\Big|^2=-\frac{d}{dt}\left(\left<\xi(t_0)\right.\left|\xi(t)\right>+c.c.\right)\vspace{3mm}\\
\dsty\hspace{2cm}=-if(t)\left(\left<\xi(t_0)\right|e^{iH_0(t-t_0)}H_1e^{-iH_0(t-t_0)}\left|\xi(t)\right>-c.c.\right).
\eer
\eeq
Integrating this expression between $t_0$ and $t_0+T$ and making use of
standard inequalities for absolute values and scalar products, we obtain:
\beq
\ber{l}
\dsty\Big|\left|\xi(t_0+T)\right>-\left|\xi(t_0)\right>\Big|^2=-i\int\limits_{t_0}^{t_0+T}dt f(t) \left(\left<e^{iH_0(t-t_0)}H_1e^{-iH_0(t-t_0)}\xi(t_0)\right.\left|\xi(t)\right>-c.c.\right)\vspace{3mm}\\
\dsty\hspace{2cm}\le 2\int\limits_{t_0}^{t_0+T}dt |f(t)| \sqrt{\left(\left|e^{iH_0(t-t_0)}H_1e^{-iH_0(t-t_0)}\xi(t_0)\right>\right)^2}\sqrt{\left(\left|\xi(t)\right>\right)^2}\vspace{3mm}\\
\dsty\hspace{2cm}\le 2\left(\mbox{max}_{[t_0,t_0+T]}|f(t)|\right)\int\limits_{0}^{T}dt \sqrt{\left<\xi(t_0)\right|e^{iH_0t}H_1^2e^{-iH_0t}\left|\xi(t_0)\right>}.
\eer
\label{bound}
\eeq
Now, assume that $f(t)$ approaches 0 at large times and consider a fixed $|\xi(t_0)\rangle\equiv |\xi_0\rangle$ (so we consider the evolution with fixed duration $T$ of the same initial state $|\xi_0\rangle$ starting at different initial times $t_0$). In this case, the first factor
in the last line becomes arbitrarily small for large $t_0$, whereas the second
factor does not depend on $t_0$. We then conclude that, for sufficiently late times,
the finite time evolution of the state vector will be approximated arbitrarily
well by $\left|\xi(t)\right>=\cnst$, i.e., by the evolution with $f(t)$ set identically
to 0.

It is then a simple corollary of the above bound that the time-dependent
matrix theory dynamics becomes approximated arbitrarily well at late times
by the flat space matrix theory, and, in particular, the supersymmetry is
asymptotically restored (with all the usual consequences, such as protection
of the flat directions of the commutator potential, and free graviton propagation).

Similar considerations can be given for the matrix string case, though they are more
involved and rely on derivations in the style of quantum adiabatic theory. We refer the reader to the original publication \cite{Craps:2010cn}.

\section{Conclusions}

We have reviewed some recent considerations of light-like singularities in string and matrix theories. In the presence of such singularities, these theories need to be supplemented with well-motivated prescriptions for singularity transition. A natural class
of such prescriptions emerges from geometrical resolutions of the light-like singularities
for which resolved space-times satisfy supergravity equations of motion. Important open
questions include analysis of the singular limit for the full non-linear evolution of matrix theories.

\section{Acknowledgments}
 
We would like to thank Frederik De Roo for collaboration on some of the material reviewed here. The research of B.C.\ has been
supported in part by the Belgian Federal Science Policy Office through the Interuniversity Attraction Pole IAP VI/11 and by FWO-Vlaanderen through project G011410N. The research of O.E.\ has been supported by grants from the Chinese Academy of Sciences and National Natural Science Foundation of China. 


\end{document}